\documentclass[runningheads,citeauthoryear]{DoklBAN}
\usepackage{epsfig,cite,graphics, graphicx} 
\usepackage{marvosym} %For astronomical symbols % % %
\usepackage[utf8]{inputenc}
\usepackage{xcolor}
\begin{document}
\title{{\color{blue} {\small Comptes rendus de l'Acad\'emie bulgare des Sciences} \\
{\small Tome 78, No 6, 2025  \\ 
$\;$ \\
$\;$ \\
$\;$ \hskip 7.4cm  ASTRONOMY}}  \\ 
$\;$ \\   $\;$ \\
PHOTOMETRY WITH THE NEW 1.5-METER TELESCOPE OF THE ROZHEN OBSERVATORY} 
\titlerunning{The 1.5m telescope of NAO Rozhen }
\author{E. Semkov$^1$,
        N. Petrov$^1$,
	M. Minev$^1$,
	M. Moyseev$^1$, 
	D. Marchev$^2$, \\
	E. Ovcharov$^3$, 
	J. Marti$^4$, 
        I. E. Dimitrova$^{1,3}$, 
	R. K. Zamanov$^1$  
	\vskip 0.3cm 
	$ \; $
	}
\authorrunning{Semkov et al.}
\papertype{{\it Presented  by N. Nedyalkov, Corresponding Member of BAS, on May 27,2025 }}
% Papertype can be "Research report", "Review", "Invited lecture", "Conference talk", 
% "Conference poster", "Lecture at scientific seminar", "Summary of dissertation",  etc.
\maketitle

\begin{abstract}
The new 1.5~m  telescope (AZ1500) is operating at the National Astronomical Observatory Rozhen, Bulgaria. 
This paper gives an overview of the telescope and  presents a snapshot of the current performance. 
Science observations are under way, and we give brief highlights from a number of programs that have been enabled. 
\footnote{{\color{blue} C. R. Acad. Bulg. Sci., 78, 797 (2025)
\\
$\;$ https://doi.org/10.7546/CRABS.2025.06.01}  }
\end{abstract}
\vskip 0.32cm 

\section{Introduction}
The exploration of the space and the nature of celestial bodies, 
their motion, origin, and evolution are among the most current and pressing topics we face today. This necessitates a continuous process of refining techniques and instruments for observation. 
We observe and study the cosmos both with ground-based instruments and those deployed on space stations, as part of a global international collaboration. 
Our National Astronomical Observatory  Rozhen is equipped 
with ground-based astronomical instruments designed for observing 
in the visible light range of the electromagnetic spectrum. 
The observatory is located on a mountain peak 
in the central part of the Rhodope Mountains. 
It is a complex that combines 
infrastructure and facilities for astronomical observations, 
scientific work and education.   

%After decades of dedicated work by several generations of astronomers, 
%in recent years we are proud to present our new observational instruments, which offer new 
% capabilities for observations and analysis of the collected data. 

%One of our most important instruments for night-sky observations 
%is the new telescope with a primary mirror diameter of 1.5 meters. 
%A new tower was constructed specifically for this purpose on the territory of NAO Rozhen, 
%with its base located at an elevation of 1750 meters above sea level. 
% The tower and its dome were built by the Italian company Gambato Astronomical Building S.r.l. 
% The dome is automated and synchronized with the telescope's pointing. 
% t is also equipped with a heating system to prevent freezing during winter operation. 
% The tower stands on a stable reinforced concrete foundation, weighing over 220 tons. 

\section{AZ1500 telescope}
The AZ1500 telescope made by ASA Astrosysteme GmbH
is a 1.5m  Ritchey-Chr\'etien  telescope equipped with quartz optics 
that guarantees  diffraction-limited performance. 
The focal length is 9~m, and the focal ratio is f/6. 
Ritchey-Chretien system is a construction variant 
that has a hyperbolic primary mirror (in our case with diameter 150~cm) 
and a hyperbolic secondary mirror (with diameter 52~cm) 
designed to eliminate optical errors. 
During the last years, the majority of the large astronomical telescopes 
are of Ritchey-Chr\'etien system, 
e.g. the Hubble Space Telescope, the ESO Very Large Telescope, the Rozhen 2m telescope. 
The mounting is Alt-Azimuth, which is a telescope mounting style
that operates around a vertical azimuth axis and a horizontal altitude axis, 
requiring variable rates on both axes to track the celestial objects. 
These mounts are good for supporting heavy loads. 
The design of the telescope results in a compact system with a weight of 5500~kg
that fits in a dome with diameter 6.0~m. 
The optical system is designed with two Nasmyth focus positions. 
First light for the telescope was achieved in the summer of 2023 [1].
For improvement of  the telescope pointing and tracking capabilities 
is created and refined a T-Point model [2].
The model that works well is with 60 stars and provides 
a pointing accuracy of about 5 arcsec. 

A new tower was constructed specifically for this telescope on the territory of NAO Rozhen. 
The tower and its dome were built by the Italian company Gambato Astronomical Building S.r.l. 
The dome is automated and synchronized with the telescope's pointing. 
The tower stands on a stable reinforced concrete foundation, weighing over 220 tons (see Fig.1). 
The coordinates are $41^0 41' 48.4''$~N,  $24^0 44' 18.4''$~E,  altitude 1750~m.  
% #       latitude = 41:41:48.4
% #       longitude = 335:15:22
%  altitude = 1750
%  335:15:41.6
%
%  \begin{figure}   
%  \includegraphics[width=3cm]{CygSNR_HaOIIIx2h.jpg}
%  \end{figure} 

\begin{figure}
\includegraphics[width=12.8cm]{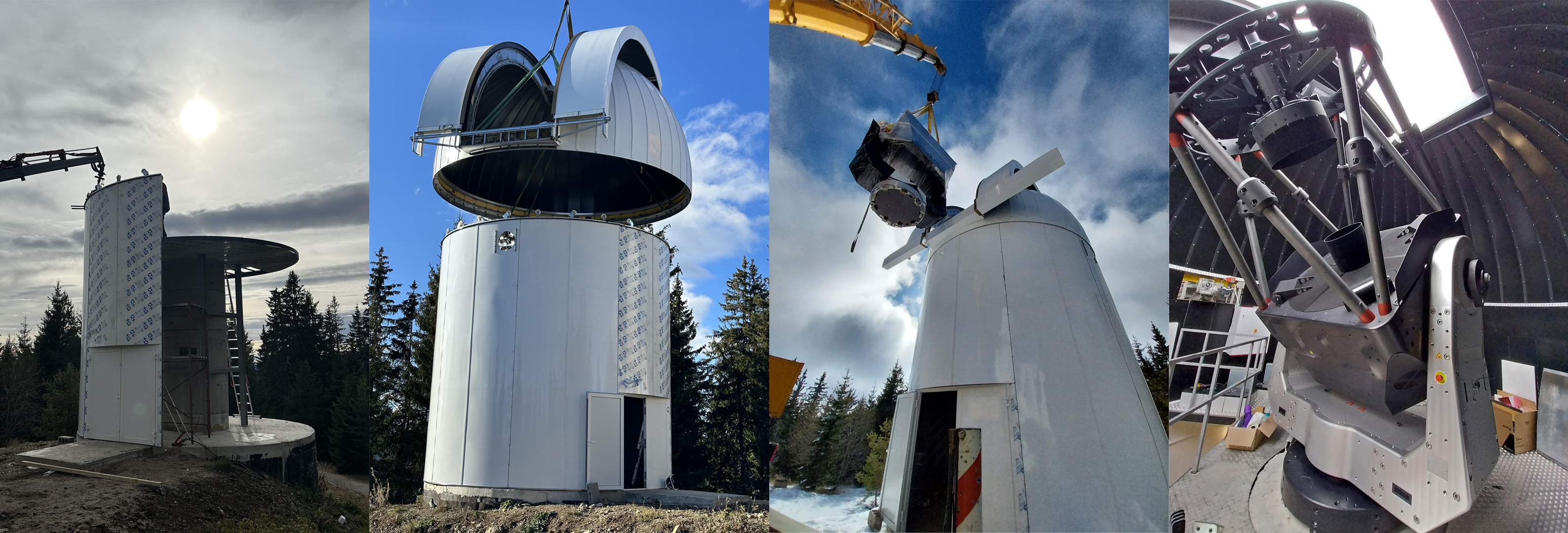}  
\caption[]{The new dome and  AZ1500 telescope. }
\end{figure}

\section{Tracking accuracy}

To give an estimation of the telescope guiding we measured the position of a few stars
on the CCD during 1 hour continuous observations.  
The shifts in pixels were transformed into arc-seconds.  
The results for a few nights are: \\

{\small
\begin{tabular}{lllllllllll}
28 Mar. 2024 &    &   UT21:19-22:19,  & \hskip 0.34cm   T CrB	    & \hskip 0.32cm  $s_1= 14.6'' $,  &   $s_2=9.8''$, \\
28 Mar. 2024 &    &   UT22:19-23:20,  & \hskip 0.34cm   T CrB	    & \hskip 0.32cm  $s_1= 1.7'' $,  &   $s_2=1.1''$, \\
15 Apr. 2024 &    &   UT01:09-02:09,  & \hskip 0.34cm   4U1954+319  & \hskip 0.32cm  $s_1= 7.7'' $,  &   $s_2=5.6''$, \\
16 July 2024 &    &   UT22:32-23:33,  & \hskip 0.34cm   T CrB	    & \hskip 0.32cm  $s_1= 0.9'' $,  &   $s_2=1.8''$, \\ 
16 July 2024 &    &   UT23:20-00:20,  & \hskip 0.34cm   T CrB	    & \hskip 0.32cm  $s_1= 8.9'' $,  &   $s_2=4.8''$, \\
3  Sep. 2024 &    &   UT18:43-19:43,  & \hskip 0.34cm   T CrB	    & \hskip 0.32cm  $s_1= 9.9'' $,  &   $s_2=4.4''$, \\
5  Sep. 2024 &    &   UT20:09-21:09,  & \hskip 0.34cm   SPICY 85657 & \hskip 0.32cm  $s_1= 1.5'' $,  &   $s_2=4.4''$, \\
\end{tabular}
}
\\

where $s_1$ is the  shift along the Right Ascension, $s_2$ is the shift 
along the  Declination,  T~CrB is a recurrent nova [3], 
4U1954+319 is a symbiotic X-ray binary consisting of neutron star and a red supergiant [4], 
SPICY~85657 is a young stellar object candidate [5].

It is visible that the average shifts are $s_1=6.5$ arcsec hour$^{-1}$ 
along  the Right Ascension  and  $s_2=4.6$ arcsec hour$^{-1}$ along the Declination.
Removing the largest value, it is equivalent to $\sqrt{s_1^2+s_2^2} = 6.5 \pm 4.1$ arcsec~hour$^{-1}$.
For a seeing of $\sim 1$~arcsec, a shift with half of the seeing 
(0.5 arsec) takes typically 4 minutes. For a single star measurements 
shifts of about 1-2 arsec would not make the photometry worse, which means that
exposures up to 10 min will provide good results.  

% Seening usually are referring to the typical FWHM's they measure in their image 

%----------------------------------------------------------------------------- 
\begin{figure}
\includegraphics[width=12.8cm]{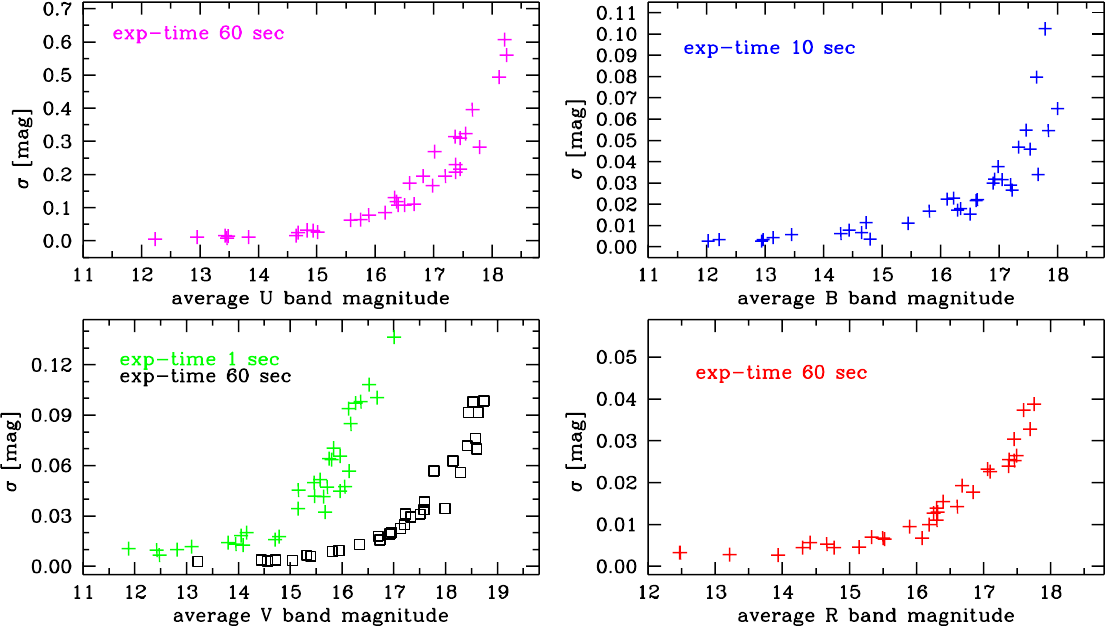}  
\caption[]{ Standard deviation ($\sigma$) versus the average magnitude 
            in U, B, V,  and R bands. 
            $\sigma$ represents the errors of the photometry.
	    The colours are:  U -- magenta, B -- blue, V -- green (exposure 1 sec), 
	    black (exposure 60 sec), R -- red. 
	    The exposure time (exp-time) is marked on each panel.
	    For more details see Sect.\ref{s.accur}.  }
\label{f.sig}
\end{figure} 
%-----------------------------------------------------------------------------

\section{The accuracy of the photometry}
\label{s.accur}

The AZ1500 telescope is equipped with a filter wheel with optical filters
in the UBVRI photometric system as well as narrow band filters.
% An example of its capability is given in Fig.~\ref{f.SNR} -- a section of the Cygnus supernova remnant.
UBVRI is a set of well-defined pass-bands in the optical range from 3300 to 8800~\AA.
Roughly speaking, the B-band covers blue, the V-band covers green, and the R-band covers red. 
As a representation of the accuracy of the observations in the different bands,  
we calculated the standard deviation of 10 consecutive measurements of $\sim 30$ stars
with different brightness: 
\begin{equation}
\sigma = \sqrt{ \frac{1}{N-1} \sum_{i=1}^{N} (m_i - \bar{m})^2 }
\end{equation}
where ${ m_1, m_2, ...., m_N}$ are the observed magnitudes, and 
$\bar{m}$ is the average value of these observations, 
$N$ is the number of observations. 
We calculated $\sigma$ for 10 exposures in UBVR bands. 
The results are plotted in Fig.~\ref{f.sig}.  
The accuracy of the observations depends on
the size of the primary mirror, the efficiency of the optics, 
filters,  CCD detector, etc.   

The  $V$~band has a mean wavelength $\lambda_{mean}= 5510$~\AA\
and  effective width $W_{eff}= 890$~\AA. 
The brightness in V-band is referred to as the apparent visual magnitude. 
As can be seen in Fig.~\ref{f.sig} for 1~sec exposure time in V band we have good 
photometry for stars in the range  $11.5 < V < 15$~mag. 
The increase of the exposure to 60 seconds, provides 
good photometry (accuracy $\pm 0.02$~mag) for stars in the range  $13 < m_V < 17$~mag.

For U-band ($\lambda_{mean} = 3605$ \AA, $W_{eff}=640$~\AA) for a 60~sec exposure, 
measurements with photometric errors $\pm 0.05$~mag can be done for stars in the range
$12 \le m_U \le 15.5$.  
For $B$-band ($\lambda_{mean}=4410$~\AA, width 960~\AA) for a 10~sec exposure  
photometry with errors $\pm 0.05$~mag can be performed in the range $12 \le m_B \le 17.5$. 
For $R$-band ($\lambda_{mean}=6580$~\AA, width 1590~\AA)  for a 60~sec exposure  
photometry with accuracy $\pm 0.02$~mag can be done for stars in the range
$12 \le m_B \le 17$. 

\vskip 0.3cm 

\section{Science programs with the telescope  }

\subsection{Accretion onto white dwarfs}

\begin{figure}
\includegraphics[width=12.8cm]{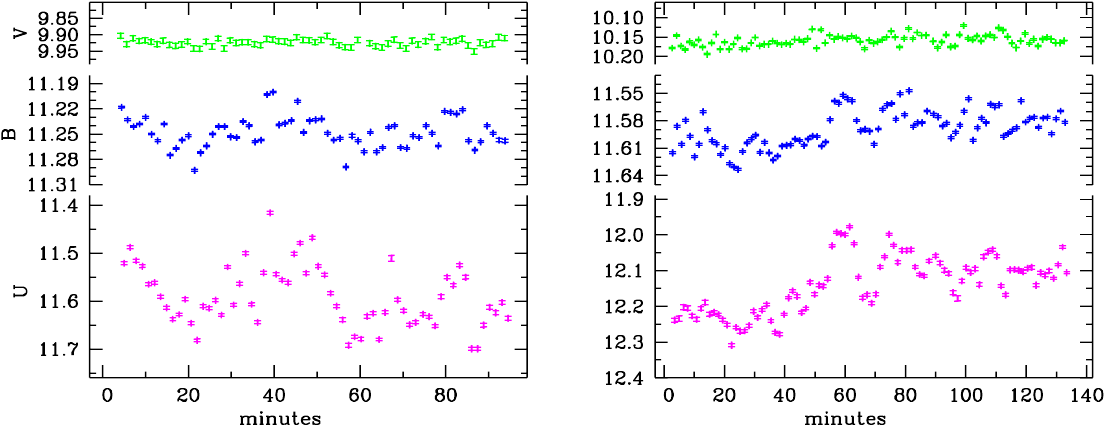}  
\caption[]{Flickering of T~CrB in U, B, V bands observed with the 1.5~m telescope. 
             The amplitude of the variability is of about 0.3 mag in $U$-band and 0.1 mag 
	     in $B-$band.   }
\end{figure}

UBV photometry of the recurrent nova T~Coronae Borealis (NOVA CrB 1866, 
NOVA CrB 1946) is already published in [6], 
where the observations with AZ1500
revealed a remarkable reddening of the system and 
decrease of the brightness and the temperature of the accretion disc in October 2023. 

Here we present two light curves of the flickering behaviour 
obtained with AZ1500 telescope. Journal of observations is presented in Table~\ref{t.2}
and the light curves are plotted in Fig.~4. 
Using the recipe  of [7] we estimate 
for the maximum,  dereddened $U-B_0$ colour 
of the flickering source $-0.98$ and $-0.49$ for 2024-04-14  and 2024-08-23, respectively.
Using the method of [8], 
we estimate for the flickering source average dereddened $U-B_0$ colour 
$-0.901$  and $-0.567$ for 2024-04-14  and 2024-08-23, respectively. 
These colour corresponds to temperature of the flickering source 
$T=13675$~K  and $T=8085$~K. The result indicates that when the brightness 
increases, the flickering  source becomes redder and its temperature decreases.
A more detailed analysis will be given in a forthcoming paper.

%--------------------------------------------------------------------------------
\begin{table*}
\caption{Observations of the intra-night variability of the recurrent nova T~CrB.  
In the first column are given date (in format YYYY-MM-DD) 
and UT of start and end of the run (in format HH:MM). In the second column are given 
the number of the exposures and the exposure time.  
The other columns give minimum, maximum and average magnitude, standard deviation, amplitude and the typical observational error. }
\centering
\begin{tabular}{lcc | cc c | c cc cc} 
\hline
  date      & band &  &  min  & max	  &  mean     &	stdev &	 ampl. &  merr  & \\
  UT        &      &  & [mag] & [mag] & [mag]     & [mag] & [mag]  & [mag]  & \\
            &   &        &	  &	      &       &        & 	& \\	    
2024-04-14  & U & 90x60s & 11.977  & 12.309 &  12.1391  & 0.079 &  0.332 &  0.005 & \\
20:41-20:52 & B & 90x10s & 11.547  & 11.633 &  11.5885  & 0.020 &  0.086 &  0.002 & \\
            & V & 90x2s  & ---   & ---    &  10.1581  & ---   &  ---   &  0.010 & \\
            &   &	&	  &	      &       &        &	& \\
2024-08-23  & U & 80x60s & 11.354  & 11.699 &  11.5804  & 0.069 &  0.345 &  0.010 & \\
18:20-20:14 & B & 80x5s  & 11.119  & 11.293 &  11.2403  & 0.033 &  0.174 &  0.005 & \\ 
            & V & 80x1s  & ---     & ---    &  9.9189   & ---   &  ---   &  0.010 & \\
            &   &        &	  &	      &       &        & 	& \\
\hline                                                           
  \end{tabular}                                                  
  \label{t.2}
\end{table*}
%--------------------------------------------------------------------------------

\subsection{Optical counterparts of high and very high energy sources}

Telescopes in the 1 to 2 meter class can play a very important support work in the golden era
of highly sensitive gamma-ray observatories. The currently existing facilities, ranging from 
the {\it Fermi} gamma-ray space telescope to the growing family 
of Imaging Atmospheric Cherenkov Telescopes (IACTs),  such as MAGIC, HESS, 
HAWC or LHAASO, will soon see their performances complemented, or  surpassed, 
by upcoming projects such as the Cherenkov Telescope Array Observatory (CTAO). 
Many new unidentified gamma-ray sources, both transient or
persistent, will certainly be discovered in a mid-term future. 
In this context, we will be facing the challenge of pinpointing their optical and/or near infrared counterparts out of the
many objects often found inside the typically $\sim 0.1^{\circ}$  gamma-ray error boxes of IACTs. This task requires not only good astrometric but also
excellent photometric capabilities, as the gamma-ray emission responsible
is often betrayed by related counterpart variability in other wavelength domains (including the optical window.)

During August 2024, the Rozhen 1.5 m telescope participated in a campaign to acquire multi-epoch photometry of the
high-mass X-ray binary candidate IRAS 18293-0941, originally identified by [9]. 
This was a good occasion to mimic the observational circumstances outlined in the previous paragraph. 
Although no short-term variability was detected, the performance of the
telescope turned out to be remarkable. 
The target was immediately outstanding in combined VRI images owing to its extremely high reddening, 
thus rendering it the most peculiar object in the region.
Both position and magnitude could be easily determined. 
Plate solutions based on a few hundreds of Gaia DR3 stars in the field of view were carried out using the IRAF software
package, in particular with its  {\tt daofind}, {\tt ccxymatch} and {\tt ccmap} tasks. An excellent 
residual root mean square, amounting to 0.01 to 0.02 arc-second, 
was achieved both in right ascension and declination for reference stars in all filters. 
All of this despite a moderate 1.5 arc-second seeing.
An accurate measurement of $0.1734 \pm  0.0001$ arc-second pixel$^{-1}$ also resulted for the CCD plate scale of the Rozhen telescope. Concerning photometry, a quick zero-point determination was conducted using a dozen reference stars in the field of view whose  magnitudes were retrieved from the AAVSO Photometric All Sky Survey (and transformed from the Sloan to the Johnson-Cousins system when needed).  The corresponding values were found to be $z_V=5.5 \pm 0.1$ mag, $z_R=4.29 \pm 0.05$ mag and $z_I=1.6 \pm 0.1$ mag, yielding 
 $V=21.9 \pm 0.3$ mag,  $R=16.4 \pm 0.1$ mag  and $I=12.6 \pm 0.1$ mag for the target IRAS source. 
Comparison with numbers in the literature strengthens the suitability of the AZ1500 telescope
for this kind of future research.

\subsection{Quasars, Novae, Star formation}

Observations of Active Galactic Nuclei, quasars [10], 
and blazars [11] with different timescales optical variability,  
in narrow and broad photometric bands are a critical aspect of modern astrophysics, providing insights into some of the most energetic  phenomena in the Universe. 
The precise  observations will  be used to apply reverberation mapping and study the
structures and physical processes around the super-massive black hole 
($10^6-10^9$~M$_\odot$) at the center.

The classical novae in the galaxy M31 have a
characteristic strong and broad H-alpha emission line. With fast and
precise observations in H-alpha filter,
novae candidates  in M31 can be confirmed or rejected.

With the AZ1500 telescope are also collected observations 
of the young stellar clusters [12], 
where can be studied star formation, Herbig's Ae/Be and T Tauri stars,  
collimated jets, nebulous filaments, embedded infrared sources 
and other young objects. 

In the future a low-resolution spectrograph might be installed in the second Nasmyth focus
for observations of transient objects, novae, supernovae,  
tidal disruption events [13],
relativistic explosions, etc. 

\vskip 0.3cm 

{\bf Conclusions:}  
The new 1.5 m telescope (AZ1500) is in operation at the National Astronomical
Observatory Rozhen, Bulgaria. 
Its design and performance are suitable for precise observations of various 
astrophysical objects -- 
from proto-stars in the star-forming regions, accreting white dwarfs, 
nova outbursts to the blazars and the quasars.

\vskip 0.3cm 

{\small {\bf Acknowledgments: }
We dedicate this paper to the memory of 
Prof. Evgeni Semkov (1961-2024), corresponding member of Bulgarian Academy of Sciences, 
who was leading the project for the telescope.  JM and RKZ acknowledge 
PID2022-136828NB-C42 grant funded by the Spanish MCIN/AEI/ 10.13039/501100011033
and "ERDF A way of making Europe". 
The telescope is a part of the National Roadmap for Scientific Infrastructure [14], 
coordinated by the Ministry of Education and Science of Bulgaria. 
	    
% \begin{thebibliography}{}
% \end{thebibliography}

 \vskip 0.4cm 
{\it 
$^1$Institute of Astronomy and National Astronomical Observatory, 
Bulgarian Academy of Sciences, Tsarigradsko Shose 72, BG-1784, Sofia, Bulgaria \\
e-mails: nip.sob@gmail.com, mminev@astro.bas.bg, mmoyseev@nao-rozhen.org
  \vskip 0.2cm 
$^2$Department of Physics and Astronomy, 
    Shumen University "Episkop Konstantin Preslavski", 115 Universitetska Str., BG-9700, 
    Shumen, Bulgaria
  \vskip 0.2cm 
$^3$Department of Astronomy, University of Sofia "Saint Kliment Ohridski", 
     5 James Bourchier, BG-1164, Sofia, Bulgaria  
 \vskip 0.2cm     
$^4$Departamento de F\'isica, Escuela Polit\'ecnica Superior de Ja\'en, 
Universidad de Ja\'en, Campus Las Lagunillas s/n, A3-420, 23071, Ja\'en, Spain
}

\end{document}